\documentclass[12pt]{aastex}
\begin{document}

\title{Observational Confirmation of a Link Between Common Envelope Binary Interaction and Planetary Nebula Shaping}
\shorttitle{Nebula-Binary Alignment and the Central Stars of NGC~6337 and Sp~1}
\author{Todd C. Hillwig\altaffilmark{1,9}}
\altaffiltext{1}{Department of Physics and Astronomy, Valparaiso University, Valparaiso, IN 46383}
\email{Todd.Hillwig@valpo.edu}
\author{David Jones\altaffilmark{2,3}}
\altaffiltext{2}{Instituto de Astrof\'isica de Canarias, E-38205 La Laguna, Tenerife, Spain}
\altaffiltext{3}{Departamento de Astrof\'isica, Universidad de La Laguna, E-38206 La Laguna, Tenerife, Spain}
\author{Orsola De Marco\altaffilmark{4}}
\altaffiltext{4}{Department of Physics and Astronomy, Macquarie University, Sydney NSW 2109, Australia }
\author{Howard E. Bond\altaffilmark{5,6,9}}
\altaffiltext{5}{Department of Astronomy \& Astrophysics, Pennsylvania State University, University Park, PA 16802 USA }
\altaffiltext{6}{Space Telescope Science Institute, 3700 San Martin Dr., Baltimore, MD 21218}
\author{Steve Margheim\altaffilmark{7}}
\altaffiltext{7}{Gemini Observatory, Southern Operations Center, Casilla 603, La Serena, Chile}
\author{David Frew\altaffilmark{8}}
\altaffiltext{8}{Department of Physics, The University of Hong Kong, Pokfulam Road, Hong Kong}
\altaffiltext{9}{Visiting astronomer, Cerro Tololo Inter-American Observatory, National Optical Astronomy Observatory, which is operated by the Association of Universities for Research in Astronomy, under contract with the National Science Foundation.}

\begin{abstract}

A current issue in the study of planetary nebulae with close binary central stars is the extent to which the binaries affect the shaping of the nebulae.  Recent studies have begun to show a high coincidence rate between nebulae with large-scale axial or point symmetries
and close binary stars.  In addition, combined binary-star and spatio-kinematic modeling of the nebulae
have demonstrated that all of the systems studied to date appear to have their central binary axis aligned with the primary
axis of the nebula.  Here we add two more systems to the list, the central stars and nebulae of NGC~6337 and Sp~1.
We show both systems to be low inclination, with their binary axis nearly aligned with our line-of-sight.  Their inclinations
match published values for the inclinations of their surrounding nebulae.  Including these two systems
with the existing sample statistically demonstrates a direct link between the central binary and the nebular
morphology.  In addition to the systems' inclinations we give ranges for other orbital parameters from binary
modeling, including updated orbital periods for the binary central stars of NGC~6337 and Sp~1.

\end{abstract}


\section{INTRODUCTION}

The shaping of planetary nebulae (PNe) has been a matter of interest for some time with the problem being approached from a
 number of different but complimentary methods \citep{kwi14}. One approach has been to identify binary central stars (CSs) 
 where the companions are close enough to have interacted in the past and determine whether the interaction might have 
 produced the observed morphologies. Since most known binary CSs are post common envelope (CE) binaries \citep[for a 
 recent review of the CE binary interaction see ][]{iva13}, it is the CE interaction that has been primarily under scrutiny.

Searches for close binary central stars of planetary nebulae (CSPNe) have been successful in discovering these systems, 
showing that approximately 10--20\% of all CSPNe appear to have a binary companion with an orbital period of less than a 
few days \citep{bon00,mis09a,dem09}. Many of those systems were discovered through photometric variability and while 
most are likely to be real binaries, additional confirmation is necessary for some of them \citep[e.g., Kn~61;][]{dem15}. Along 
with studies confirming the binarity of several of these systems \citep[e.g.][]{shi08,hil15a,hil16}, discoveries of additional 
close binary CSPNe are helping us to better understand the nature of these systems. In addition, studies of the CS can be 
linked to kinematic studies of the nebulae to determine whether a causal link exists between the interaction and the 
nebular morphology and kinematics.

Surprisingly, while there are several suggestions that the CE interaction is the cause for the shape of post-CE PNe 
\citep{mor81,bon90,zij07,dem09,mis09b} there has never been a quantification of the link. The reason has been a lack of data. 
Binary modeling of known and newly discovered binary CSPNe along with spatio- kinematic modeling of the PNe have 
shown that of the systems studied, all seem to show an alignment between the central binary axis and the primary 
geometrical axis of the PN. Including the two systems we provide values for in this paper, there are now eight known PNe 
with binary CSs for which both inclinations are known. Using these data we are here, for the first time, demonstrating a 
correlation between post-CE CSs and their PN shapes, which we argue below implies causation.

\section{Observations And Reductions}

\subsection{Photometry}

We utilize the orbit-resolved $B$, $V$, and $R$ photometry from \citet*{hil10} for the CS of NGC~6337. 
As described in \citet{hil10}, the photometry shows clear and consistent, nearly sinusoidal, variability of nearly
identical amplitudes in all three filters.  The photometry is presented in differential magnitudes via single star
differential photometry with two additional comparison stars used to confirm that our first comparison star is
not photometrically variable.

Photometric variability of the nucleus of Sp~1 was originally discovered in 1988 by H.~E.~Bond.  It was chosen for photometric 
monitoring because \citet{men88} had reported an emission-line spectrum similar to that of the close binary central star of 
HFG~1, likely due to reprocessed radiation on the facing hemisphere of a companion. Observations in $B, V, R,$ and $I$ 
were made by H.~E.~Bond on the CTIO 0.9-m telescope during five observing runs in 1988-90, and a preliminary photometric 
period of 2.91 days was reported in \citet{bon90}. The full data have not been published previously but are presented here.

Also included are new data from a number of telescopes: $I$-band data from the 0.6-m SARA telescope at CTIO (SARA-CT) in
2010 and 2011, $V$, $R$, and $I$ band photometry from the CTIO 0.9-m and 1.3-m telescopes in 2009, and more $I$-band data
from the South African Astronomical Observatory in 2010.

The data show very consistent periodic
behavior over the entire 23 year span of observations.  However, there is clear scatter around the nearly sinusoidal
average that is larger than would be consistent with the uncertainties in the photometry.  We discuss the implications
and possible causes of this scatter below.

The apparent magnitudes for the CS of Sp~1 were calibrated using
secondary standard stars in the Sp~1 field.  Absolute photometry of the secondary standards was performed on a
photometric night using Landolt standard stars \citep{lan92}.

Tables~\ref{dmag6337} and~\ref{magsp1tab} give the differential and apparent magnitudes
for the CSs of NGC~6337 and Sp1~1 respectively (see the electronic version for the full tables).

\begin{center}
\begin{deluxetable}{lcccccccc}
\tablewidth{0pc}
\tablecolumns{9}
\tablecaption{Differential Magnitudes of the Central Star of NGC~6337
\label{dmag6337}}
\tablehead{
\colhead {HJD}        & \colhead {$B_{diff}$} & \colhead {$\sigma_B$}  & 
\colhead {HJD}        & \colhead {$V_{diff}$} & \colhead {$\sigma_V$}  & 
\colhead {HJD}        & \colhead {$R_{diff}$} & \colhead {$\sigma_R$}  \\
\colhead {(2450000+)} & \colhead {(mag)}  & \colhead {(mag)}  &
\colhead {(2450000+)} & \colhead {(mag)} & \colhead {(mag)}  &
\colhead {(2450000+)} & \colhead {(mag)} & \colhead {(mag)}  }
\startdata
3477.79437 & 2.372 &  0.003 &  3477.79884 & 2.156 & 0.004 & 3477.80332 & 1.966 &  0.003   \\
3477.80804 & 2.289  & 0.003 &  3477.81252 & 2.052 & 0.003 & 3477.81700 & 1.879 &  0.003   \\
3477.82262 & 2.195  & 0.003 &  3477.82710 & 1.981 & 0.004 & 3477.83157 & 1.838 &  0.003    

\enddata
\tablecomments{Table 1 is published in its entirety in the machine-readable format.
      A portion is shown here for guidance regarding its form and content.}
\end{deluxetable}
\end{center}

\begin{center}
\begin{deluxetable}{lccccccccccc}
\rotate
\tablewidth{0pc}
\tablecolumns{9}
\tablecaption{Apparent Magnitudes of the Central Star of Sp~1
\label{magsp1tab}}
\tablehead{
\colhead {HJD}        & \colhead {$B$} & \colhead {$\sigma_B$}  & 
\colhead {HJD}        & \colhead {$V$} & \colhead {$\sigma_V$}  & 
\colhead {HJD}        & \colhead {$R$} & \colhead {$\sigma_R$}  &
\colhead {HJD}        & \colhead {$I$} & \colhead {$\sigma_I$}  \\
\colhead {(2400000+)} & \colhead {(mag)}  & \colhead {(mag)}  &
\colhead {(2400000+)} & \colhead {(mag)} & \colhead {(mag)}  &
\colhead {(2400000+)} & \colhead {(mag)} & \colhead {(mag)}  &
\colhead {(2400000+)} & \colhead {(mag)} & \colhead {(mag)}  }
\startdata
47223.7965 & 14.432 & 0.003 & 47223.8060 & 13.953 & 0.003 & 47223.8069 & 13.656 & 0.003 & 47223.8079 & 13.289 &    0.0040 \\
47223.8008 & 14.436 & 0.003 & 47223.7977 & 13.948 & 0.003 & 47223.7988 & 13.650 & 0.003 & 47223.7998 & 13.291 &    0.0040 \\
47223.8050 & 14.431 & 0.003 & 47223.8018 & 13.956 & 0.003 & 47223.8028 & 13.651 & 0.003 & 47223.8040 & 13.296 &    0.0040 

\enddata
\tablecomments{Table 2 is published in its entirety in the machine-readable format.
      A portion is shown here for guidance regarding its form and content.}
\end{deluxetable}
\end{center}

\subsection{Spectroscopy}

We also obtained orbit-resolved spectroscopy for both CSs using the Gemini South telescope and the GMOS-S instrument in long-slit mode.  The B1200 grating was used with a 0.75 arc second wide slit with 2x2 binning and a central wavelength of 4750 \AA~resulting in a wavelength range from 4000--5460 \AA. The resolution of the B1200 grating is 3744, giving a $\Delta\lambda$ at our central wavelength of
1.27 \AA.  The resulting spectra have a pixel size in the spectral direction of 0.47 \AA.  
We obtained ten pairs of 600 second exposures for NGC~6337 spread over the orbital
period and for Sp~1 we obtained 15 pairs of 480 second exposures.

The spectra of the CS of NGC~6337 show clear emission
lines due to an irradiated secondary as well as radial velocity variability with the photometric period.  Also visible is broad
hydrogen emission similar to those seen in HFG~1 \citep{ext05} and Abell~65 \citep{hil15a}.  In this case as well,
the broad H emission with narrow absorption core corresponds to the irradiated companion.

The spectra of the CS of Sp~1 show emission lines from an irradiated cool companion as well as absorption lines of H and \ion{He}{2}.  Both sets of features vary in radial velocity with the photometric period.

All spectra were reduced with the GEMINI package in IRAF.  Wavelength calibration was performed using CuAr arc spectra.  
Radial velocity values for each spectrum were found using line centers.  For emission lines and weak absorption lines line
centers were found by Gaussian line fitting while for strong absorption features we used Voigt profile fitting.  A resulting
radial velocity for each spectrum used the average for a set of lines.  The lines used for each system are described below.
We also used cross-correlation fitting (CCF) to find radial velocities.  The results from the two methods agreed to within uncertainties,
however because of the irradiation effect in both systems causing variations in line strength, the CCF results typically had
larger uncertainties than simple centroid averaging of multiple lines.

For the initial visits to both targets, arc spectra were obtained at the beginning of the night rather than consecutive with
the science spectra.  In these spectra we found overall
 shifts of as much as 15 km s$^{-1}$ in the average value of nebular line positions (which should remain stationary).  
 Based on these results, all subsequent visits
 included arc spectra taken immediately before or after the science spectra.  Using these arc spectra, the standard deviation
 of random spectral shifts between spectra were reduced to $<2$ km s$^{-1}$ for the average positions of nebular lines.

Despite the average of all available nebular lines from one spectrum to another being consistent to within
a few km s$^{-1}$, comparing {\it individual lines} for a given a spectrum we find line-center differences
of up to 15 km s$^{-1}$, with a standard deviation of up to 5 km s$^{-1}$.  Given the spectral resolution and
$\Delta\lambda$ of our spectra, these calibrations correspond to about one-tenth of a resolution element which we find to be an
acceptable calibration.  However, the availability of the nebular lines allows us to {\it internally} calibrate each science spectrum.  This
includes the early spectra which did not have arc spectra taken consecutively.  We find an average and standard deviation
for all spectral lines across all spectra (removing those with no consecutive arc spectra) for each system.  We then use
the difference of individual spectral lines in each spectrum from that average to create an internal wavelength calibration for the stellar spectra.
Such an approach has been used successfully in the past \citep[e.g.][]{jon15}.

For Sp~1 we find from the nebular lines a PN systemic velocity of $\gamma_{PN} = -27\pm2$ km s$^{-1}$ and for
NGC~6337 a PN systemic velocity of $-69\pm5$ km s$^{-1}$.  We described these results in more context below.

For the CS and companion in Sp~1 the total radial velocity variation is only about 20 km s$^{-1}$, the results were then very
sensitive to the calibration variations mentioned above.  However, in Figure \ref{sp1rv} we show the double-lined radial velocity
curve for the central binary in Sp~1, demonstrating that our final calibration is good enough to detect and measure the radial velocity
variations of both components.  The narrow emission lines provide the most precise curve of the two, while the 
broader absorption lines have correspondingly larger uncertainties and scatter.  We find that the hydrogen absorption
lines are not consistent throughout the orbit, often having what appears to be a blue-shifted absorption component
of the line, though the overall shape varies as well.  We used multiple-component line fitting in an attempt to
improve the radial velocity curve.  However we find that even with the improved fits, using only the hydrogen lines results
essentially in a scatter plot in radial velocity.  Thus the CS curve in Figure \ref{sp1rv} uses only the \ion{He}{2} absorption lines
at $\lambda\lambda$ 4541, 4686, and 5411 \AA.

For the CS of NGC~6337, we have only the emission lines and as mentioned, the radial velocity amplitude is much larger.  
Therefore the resulting radial velocity curve of the cool companion in that system (Figure \ref{ngc6337rv}) has considerably better 
precision relative to amplitude.

\begin{figure}[p]
\begin{center}
\includegraphics[angle=-90, width=6in]{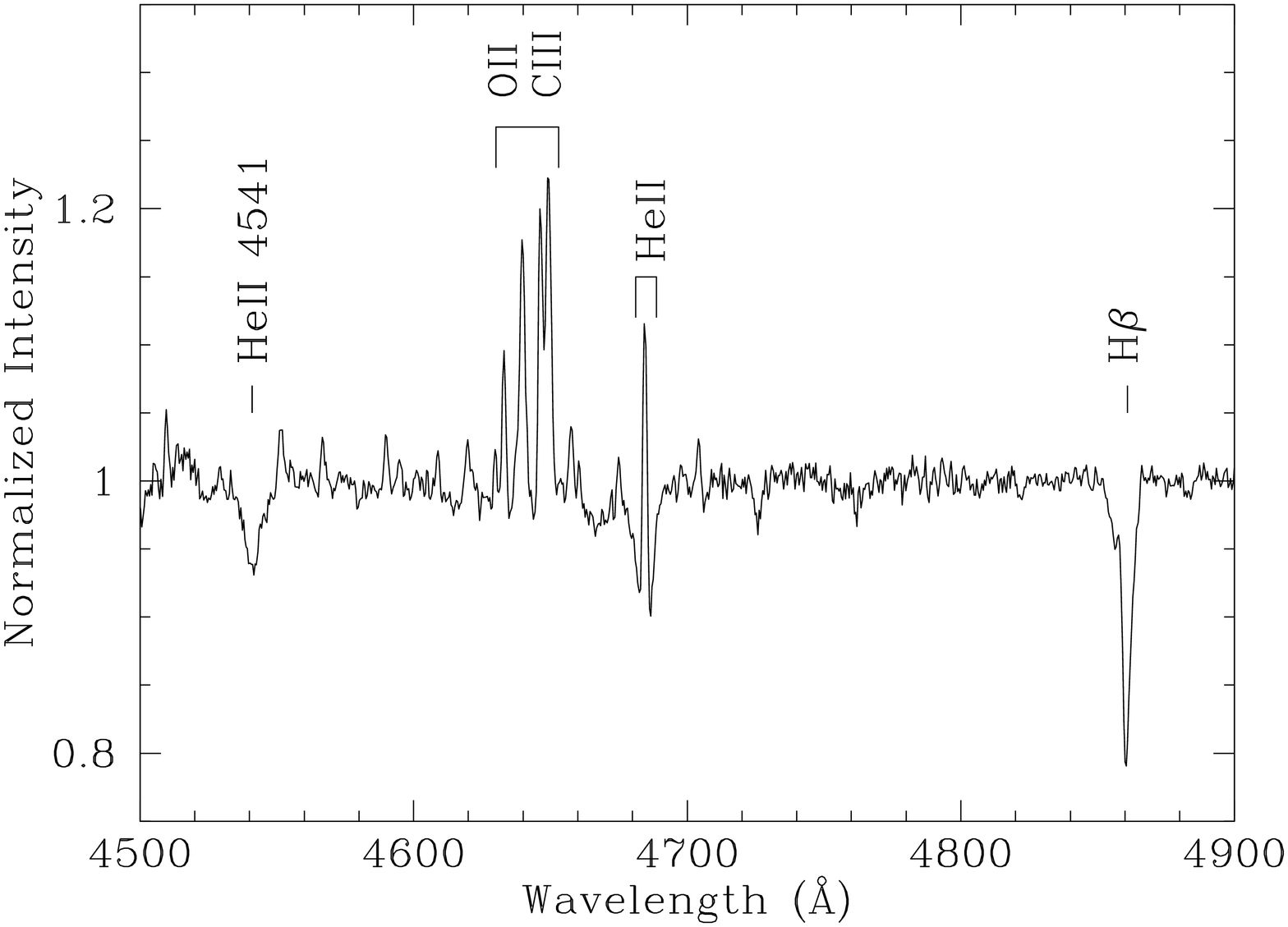}
\end{center}
\caption[Sp~1 Spectrum] {An example of the continuum normalized spectrum of the CS of Sp~1
showing both emission features from the irradiated secondary and absorption features from the hot CS.  \label{sp1spec}}
\end{figure}

\begin{figure}[p]
\begin{center}
\includegraphics[angle=-90, width=6in]{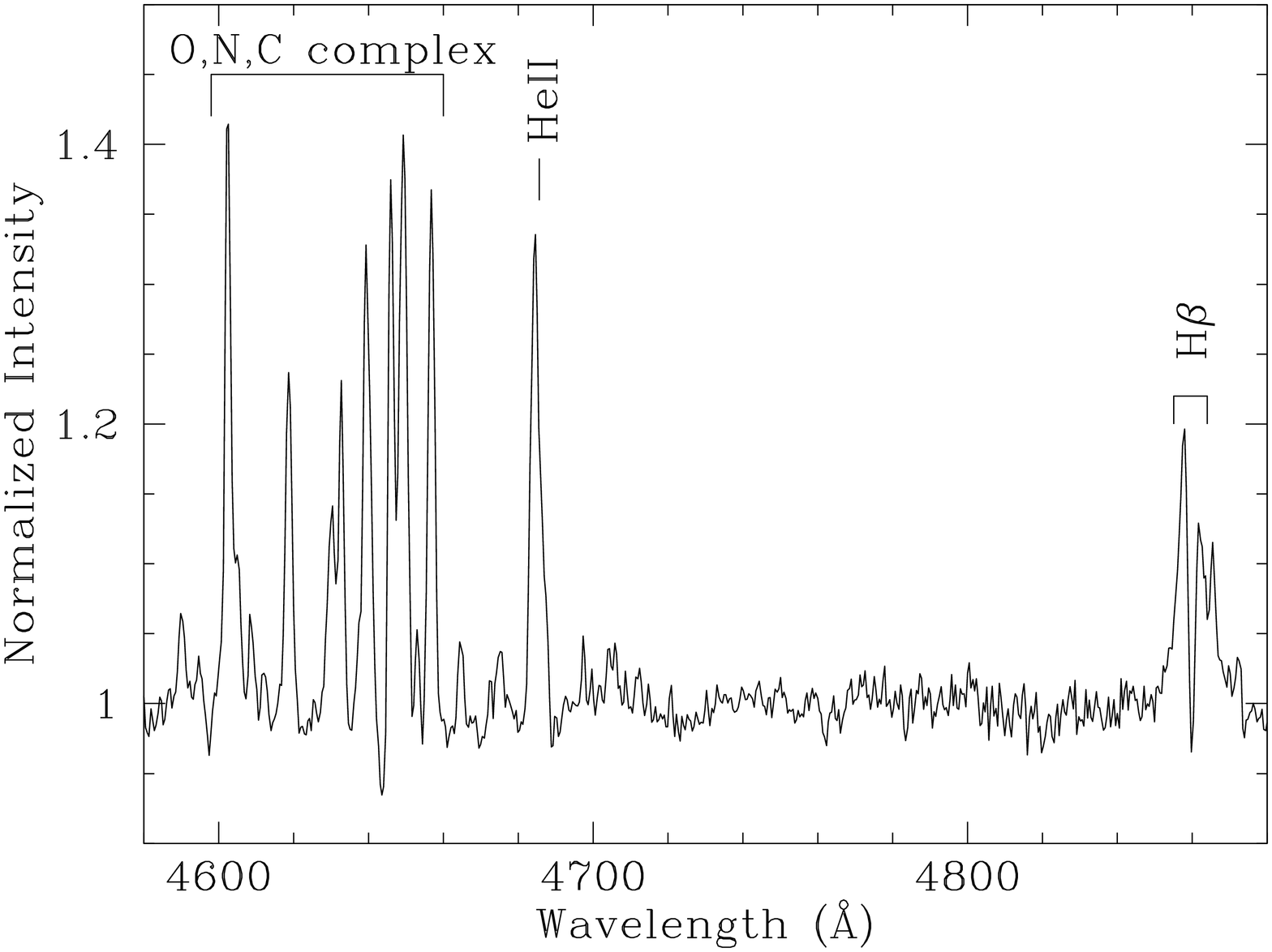}
\end{center}
\caption[NGC~6337 Spectrum] {An example of the continuum normalized spectrum of the CS of NGC~6337
showing emission features from the irradiated secondary, including the broad H$\beta$ feature.  \label{ngc6337spec}}
\end{figure}

\section{NGC~6337}

\subsection{Overview}

The CS of NGC~6337 was found to be a photometric variable consistent with a close binary system by
\citet{hil04}, with preliminary binary system modeling performed later \citep{hil10}.  The reported period
was 0.1734742 days though with some uncertainty due to possible aliasing.  The range of possible
models suggested a low to intermediate inclination binary system with a late M star companion and a
CS that had nearly reached the WD cooling track.  
\citet{wei15} describe the emission-line spectrum of NGC~6337's
CS, which is consistent with the finding that the system harbored an irradiated binary.

\begin{figure}[p]
\begin{center}
\includegraphics[width=6in]{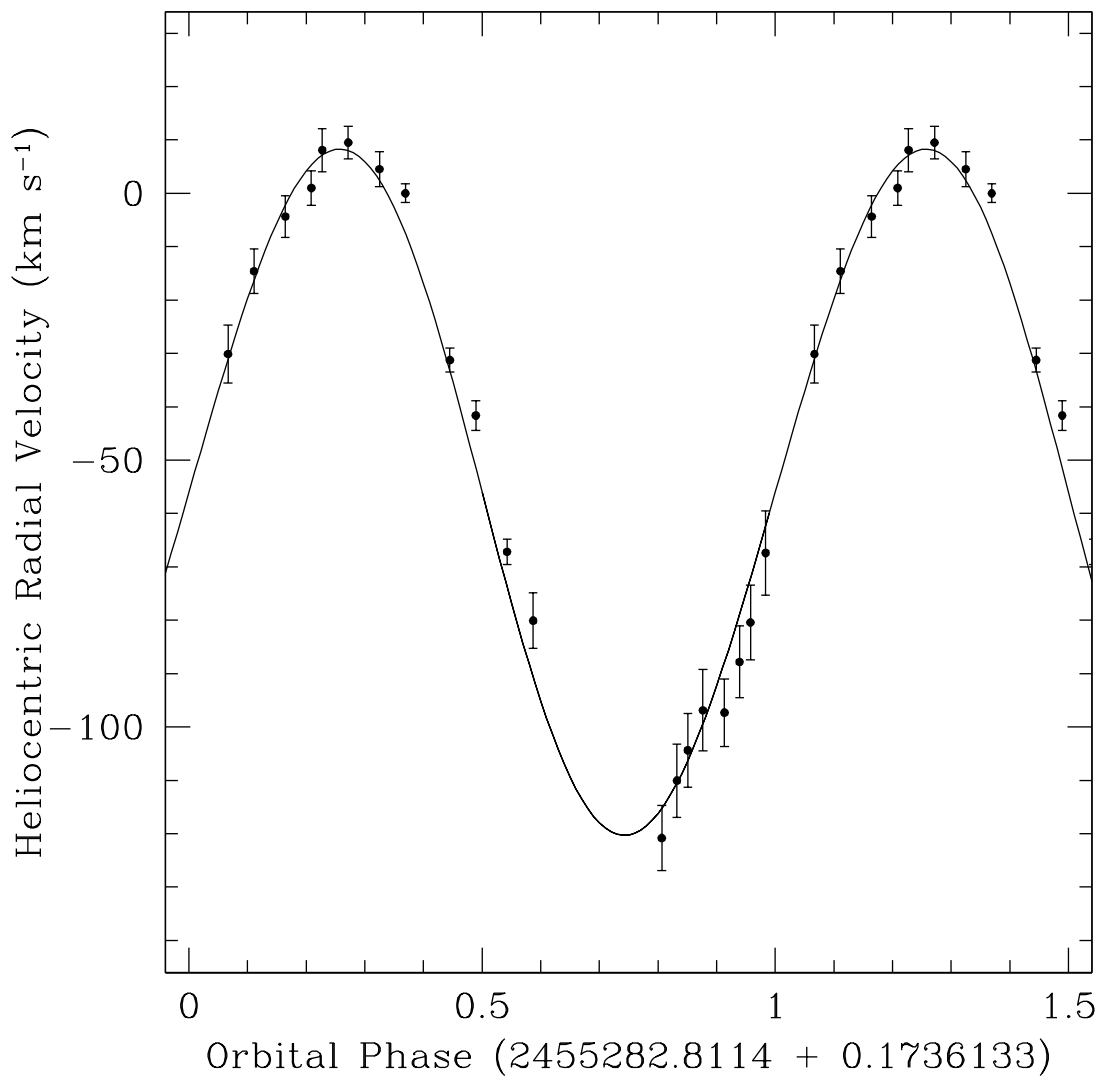}
\end{center}
\caption[NGC~6337 Phase-Folded Radial-Velocity Curve] {The phase-folded radial-velocity curve
of NGC~6337 for the period given in the ephemeris (Equation \ref{eq2}).  
The fit line is from the Wilson-Devinney model \citep{wil71,wil90} described in \S 3.1.2.  We find a system
velocity of $\gamma=-54\pm3$ km s$^{-1}$ with a radial-velocity amplitude for the secondary of
$K_{CS}=63\pm2$ km s$^{-1}$.
\label{ngc6337rv}}
\end{figure}

The radial velocity curve resulting from our data shows the variation of the emission lines from the heated 
secondary star (Figure \ref{ngc6337rv}), as described previously.  The period of the radial velocity variations
matches those of the photometric variations, confirming that this is indeed a binary system exhibiting an
irradiation effect.  However, the radial velocity data, combined with the photometry provides a longer baseline,
allowing us to improve the period determination.  We find that the primary period reported in \citet{hil10} was in
fact an alias period, one that was removed with the additional data.  The true period is very close to the
previously reported value.  The improved ephemeris in HJD is
\begin{equation}
T= 2455282.8114(5)+0.1736133(5)E. 
\label{eq2}
\end{equation}
We use this improved period in our binary modeling, though in terms of the absolute physical parameters,
the effect of the adjusted period is smaller than the uncertainties in the resulting parameters.

Nebular morphology modeling \citep{gar09} suggests
a nebular inclination $\lesssim 10^\circ$ based on the nearly circular
appearance of the equatorial ring (thus nearly face-on) and low measured expansion velocities for the bright ring.
The color image of NGC~6337 from \citet{hil10} is shown in Figure \ref{PNimages}.
We find from our images that the ring is not
sharply delineated.  Thus, assuming the equatorial ring
is uniform in radius, inclinations as high as $\sim20^\circ$ still produce a nearly circular appearance.  
An inclination of $20^\circ$ would only produce a 6\% difference
between the short and long axes of the projected ellipse.  This corresponds to less than one-fifth of the width of the
emission in the ring.  In this case it would be difficult to detect such a small difference.  
Additionally, if the narrow waist is not thin, but has some measurable
vertical extent, as suggested by \citet{cor00}, then the axis ratios can become more complicated.    
In terms of the measured projected expansion
velocities, a difference in inclination between $10^\circ$ and $20^\circ$ only results in a factor of two difference
in the true expansion velocities.  Given that the measured value quoted by \citet{gar09} is uncertain
to a factor of two, and that the ring shows internal motions up to 50 km s$^{-1}$ \citep{cor00}, this
range in inclinations does not seem improbable.

Our long-slit spectra, taken at a position angle (PA) of $90^\circ$, provide radial velocities of the
bright ring on either side of the CS.  We find a difference in the radial expansion velocities of the ring
at this position angle to be $\Delta V_{r}(\mathrm{exp,PA}=90)=9.3\pm0.5$ km s$^{-1}$, giving a projected expansion
velocity of $V_{r}(\mathrm{exp,PA}=90)\approx4.6$ km s$^{-1}$.  Assuming the nebular symmetry axis has 
PA $\approx-45^\circ$ from \citet{gar09}, and a tilted, symmetric ring or toroid,
our measured value should be $\cos 45^\circ = 0.707$ times the maximum radial
expansion velocity of the ring measured at PA $=45^\circ$, giving $V_{r}(\mathrm{exp,PA}=45)\approx6.5$ km s$^{-1}$.
If the ring is tilted by $20^\circ$, the resulting deprojected expansion velocity would be about 19 km s$^{-1}$, which
is consistent with the expected speed of slow winds and the expansion velocities of equatorial rings in these systems.

If we also look at the lower end of the inclination range for the NGC~6337 nebula, the faint but visible lobes shown by
\citet{cor00} and \citet{gar09}, showing what they identify as caps and jets, demonstrate that an inclination near zero
is very unlikely.  These features show point-symmetry around the central star with the caps and jets located at
slightly different PAs.  Neither lobe shows much lateral extent from the $\approx 45^\circ$ PA of the assumed major axis,
suggesting that the lobes are not significantly wider than the bright ring.  If we assume that the lobes and ring are
cylindrical (have the same lateral extent from the symmetry axis), then using the angular extent of the lobes on the
sky we can determine a relationship between the nebular inclination and the axial ratio (the height of the cylinder
relative to the diameter or half-height, $h$, relative to the radius $r$) of the nebula for small inclinations, 
$$h/r = \frac{1}{\sin i}(\xi+\frac{\sin^2 i}{\cos i})\approx \frac{\xi}{\sin i}$$
where $\xi$ is the fractional extent of the lobes outside of the ring relative to the radius of the ring.  From our CCD images
of NGC~6337 we find a somewhat conservative value $\xi\approx 0.6$.  In other words, the lobes appear to extend
approximately 0.6 ring radii outside of the bright ring.  Using this relationship for $i=5^\circ$ gives $h/r\approx 7$.  Even for
extreme nebulae, such as Hen~3-401, the axial ratio is around 5.  For $i=20^\circ$ we find $h/r\approx 2$ which is a
reasonable value for a young PN and perhaps nearing the lower expected limit for similar systems.

Based on these arguments we use a range of $5^\circ$ -- $20^\circ$ for the inclination of the PN.

System velocities from the literature fall very close to -70 km s$^{-1}$ \citep*{mea88,cor00,gar09}.
The average heliocentric velocity for our measurements of the ring at PA $=90^\circ$ is $-69\pm5$ km s$^{-1}$,
consistent with the previous values.  However, our system velocity for the binary, determined from the radial velocity
curve of the companion (Figure \ref{ngc6337rv}) is $\gamma=-54\pm3$ km s$^{-1}$, which is measurably different
than the nebular velocity.  It is unclear why the binary system velocity seems to differ from the nebular velocity.
There are several possible reasons including a triple system, some kind of kick to the binary system relative to the nebula,
deceleration of the nebula due to interaction with the surrounding ISM, or a brightness center of the nebula that is offset
from the geometric center \citep[e.g. see the case of SuWt~2 in][]{jon10}.  We do not have clear evidence for any
of these options so at present the reason remains unknown.  Further discussion of possible causes of off-center CSPNe
are given by \citet{sok98} and \citet{sok15}.

The previously published distances to NGC~6337 described in \citet{hil10} range from 0.86--1.7 kpc and are consistent
with the recent mean value from \citet{fre16}, $d=1.45\pm0.43$ kpc.  \citet{fre16} also give $E(B-V)=0.60\pm0.14$,
resulting in $A_V=1.86$.
We use the range in distance as a 
consistency check with our binary modeling, as described in \S 3.2.

\begin{figure}[p]
\begin{center}
\includegraphics[angle=0,width=6in]{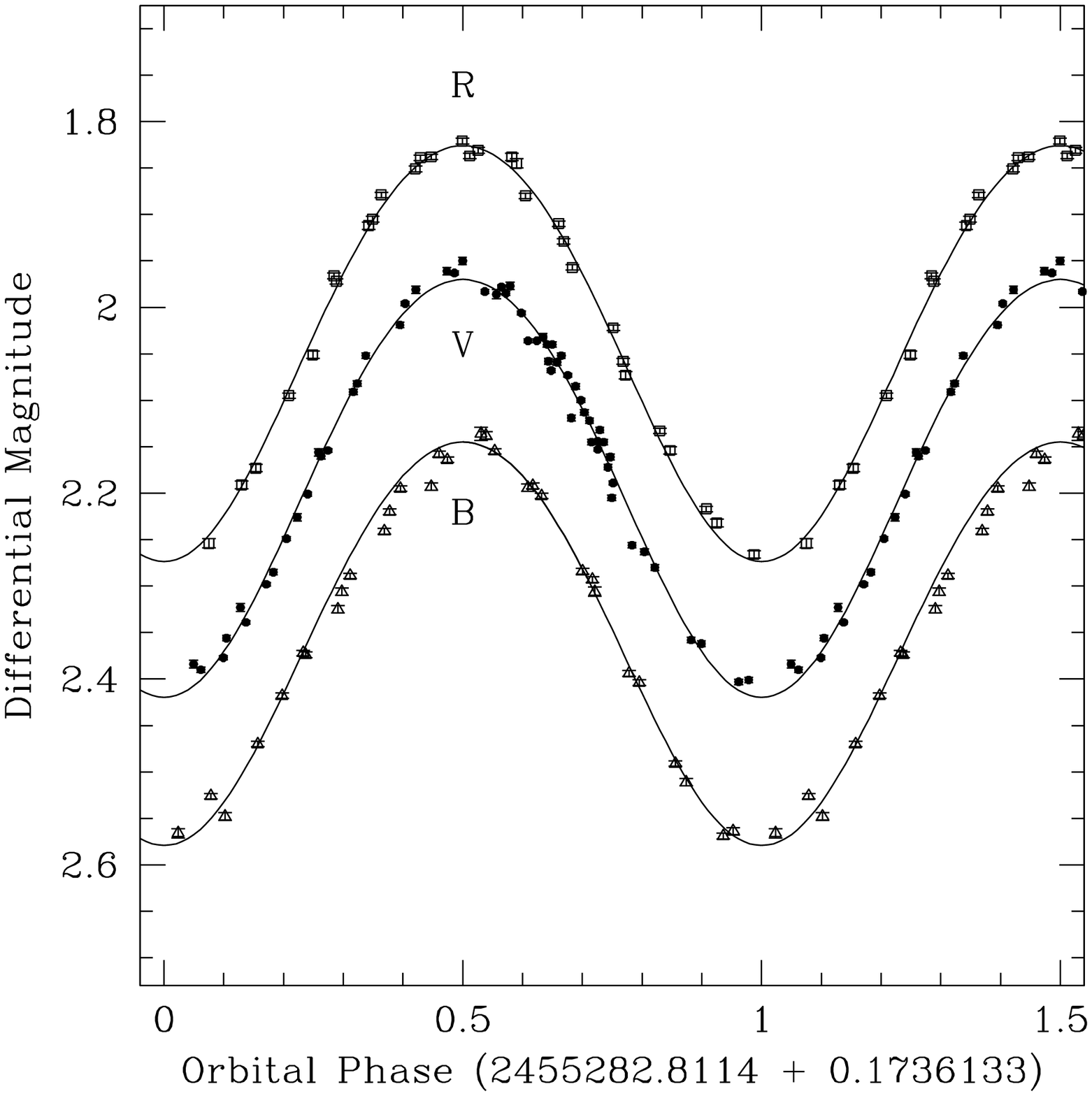}
\end{center}
\caption[NGC~6337 Phase-Folded BVR Light Curve] {The phase-folded $B$, $V$, and
$R$-band light curves of the CS of NGC~6337 for the period given in the ephemeris
(Equation \ref{eq2}).  The curves have been vertically displaced from one another for clarity.
The fit lines are from the Wilson-Devinney model discussed in \S 3.1.2.
\label{ngc6337phot}}
\end{figure}

\subsection{NGC~6337: Modeling the light and radial velocity curves}

Given our previous modeling \citep{hil10} which we used as a starting point for this work, the addition
of a single-lined radial velocity curve allowed us to provide much stronger limits on the possible
system parameters.  The amplitude of the companion's radial velocity curve was found to be
$K_2=63\pm2$ km/s.  However, since this was found using the emission lines in the spectrum, it
represents the amplitude of the center of light of the irradiated hemisphere rather than the center of
mass of the star.  Because the Wilson-Devinney code uses the center of light to produce its radial
velocity points, and because we expect the irradiated hemisphere to dominate the light from the
companion, the effect is corrected, at least to first order, in our modeling.

For our models we set the mass of the CS to be $M_{CS}=0.56$ M$_\odot$ based on the value
from \citet{gor97}.  We discuss below the consistency of this value with our resulting parameter sets.

The results of our modeling are given in Table \ref{CSmodels} and are shown as solid lines
in Figures \ref{ngc6337rv} and \ref{ngc6337phot}.  The well-determined value for
the companion's radial velocity amplitude (along with the set value for $M_{CS}$) provide a narrow
range of system inclinations from our binary system modeling.
If we allow the CS mass to vary slightly, then the inclination range expands with it, 
however for a few tenths of a solar mass uncertainty in $M_{CS}$ the inclination range expands by less than
one degree.  The resulting range of $17^\circ\leq i\leq 23^\circ$ overlaps with our adopted nebular inclination range.
We also note that the model radial velocities are determined for the center of light, rather than
the center of mass of the star.  Since the radial velocity curve was determined from irradiation emission lines,
and the irradiated hemisphere will dominate the light from the star, the model and data should be
measuring the same thing.

Our parameter values for the secondary, $M_2$, $R_2$ and $T_2$ fall across roughly the same range as our previous 
results \citep{hil10}, with the range of possible companion masses, $M_2$, reaching slightly lower values.
We have produced more precise results for the radius and temperature of the CS.  We then find the
resulting bolometric luminosity of the CS to be 467 L$_\odot$ $\lesssim L \lesssim 1100$ L$_\odot$.
Comparing these values for the CS with the post-AGB models of \citet{sch83} we find that the results
fall close to the evolutionary track of a 0.565 M$_\odot$ CS with an age of between 20,000 and 24,000 years.
Thus our assumption of the CS mass is consistent with our results.

Using VPHAS data \citep{dre14} for the CS of NGC~6337 with u = 14.908 +/- 0.004, (MJD = 56566.00580)
u = 14.909 +/- 0.004 (MJD = 56566.00805),
g = 16.206 +/- 0.003 (MJD = 56566.01689), 
g = 16.176 +/- 0.003 (MJD = 56566.01931), 
r2 = 15.805 +/- 0.004 (MJD = 56566.02531), and
r2 = 15.800 +/- 0.004 (MJD = 56566.02611)
with VPHAS+ filter profiles, we can interpolate to get $V = 16.00 \pm 0.04$ at MJD 56566.02, which corresponds
to orbital phase $0.071\pm0.001$.  Using our $V$ band light curve we then find a minimum light magnitude of
$16.02\pm0.04$.
Using this as the apparent magnitude, the interstellar
absorption value described above, and our modeled parameters for the binary, we can calculate a distance to
NGC~6337.  In our models, we find that because of the low inclination, even at minimum light we are seeing
a large portion of the irradiated hemisphere of the companion.  Therefore, at minimum light the companion
contributes roughly 30\% as much light as the CS in $V$.  Using the total brightness at minimum light we find 
$d=$0.90--1.3 kpc, which overlaps well with the distance estimates from the literature.

As with nearly all of the previous companion stars in these close binary systems, the companion here is
over-luminous compared to main sequence stars of its mass.  The radius in this case is similar to, or slightly
larger than for an equivalent mass MS star, and the temperature is measurably higher.

\begin{center}
\begin{deluxetable}{lcc}
\tablewidth{0pc}
\tablecolumns{2}
\tablecaption{Best-fit Model Physical Parameters for the CSs of NGC 6337 and Sp 1
\label{CSmodels}}
\tablehead{
\colhead {Parameter}       & \colhead{NGC 6337} & \colhead{Sp 1}}
\startdata
$M_{CS}$ ($M_\odot$) 		& 0.56  	         & 0.52--0.60 \\
$M_2$ ($M_\odot$) 			& 0.14--0.35  	 & 0.52--0.90 \\
$T_{CS}$ ($\times10^3$ K) 	& $115\pm5$  	 & $80\pm10$ \\
$T_2$ ($\times10^3$ K) 		& $4.5\pm0.5$ 	 & 3.5--4.6  \\
$R_{CS}$ (Pole, $R_\odot$) 	& 0.045--0.085  & 0.20--0.35 \\
$R_2$ ($R_\odot$) 			& 0.30--0.42  	 & 1.05--1.6 \\
$i$ ($^\circ$) 				& 17--23  		 & 7--11 \\
\enddata
\end{deluxetable}
\end{center}

\section{Sp~1}

\subsection{Overview}

The PN Shapley~1 was discovered by \citet{sha36} who described it as a ``fine ring'' nebula with total diameter 72 arc seconds
and width of the ring of 15 arc seconds.  Figure \ref{PNimages} shows a color composite image of Sp~1 using $u,V,$ and $R$
images.  \citet{sab86} gives a distance for the PN of 1.5 kpc and \citet{fre16} give a mean statistical
distance, $d_{mean}=1.46\pm0.43$ kpc using their surface brightness method, and an optically-thin distance, 
$d_{thin}=1.19\pm0.24$ kpc.  \citet{fre16} also give an interstellar reddening value 
$E(B-V)=0.56\pm0.13$.

%
The CSPN of Sp~1 was discovered to be a periodic photometric variable, and likely binary, by \citet{bon90}, who give
a period of 2.91 days.   Because Sp~1 appears to be a nearly perfect circular ring, it was anticipated that the CS, if it 
was a binary, was seen nearly pole-on. In fact, the photometric amplitude was found to be small, about 0.1 mag in $B$ 
\citep{bon90}.  The orbital
period and photometric amplitude were confirmed by \citet{bod12} who also published $V$, $R$, and $I$ photometry
as well, showing increasing amplitudes at redder wavelengths, as expected for a cool irradiated companion in such systems
\citep{dem08}.  \citet{bod12} also performed preliminary binary modeling of the photometry using values from the literature
for $T_{CS}$, $R_{CS}$, and $M_{CS}$.  They also limited possible companions to MS stars.  Even though
most cool companions in short period CSPN binaries are larger and hotter than their MS counterparts, this limitation
allowed preliminary modeling with only light curves to produce converging parameter sets.  Their resulting ranges
for the physical parameters of the companion and the system inclination, based on their imposed modeling limits were
$1.16\lesssim M_2\lesssim 1.25$ M$_\odot$, $1.12\lesssim R_2\lesssim 1.20$ R$_\odot$, 
$5900\lesssim T_2\lesssim 6100$ K, and $15^\circ\lesssim M_2\lesssim 25^\circ$ M$_\odot$, 
We compare our results, using
fewer initial assumptions and including radial velocity curves, with theirs in \S 3.2.3.

Spatiokinematic modeling of the nebula was performed by \citet{jon12}, who showed that the nebula appears to be nearly
face-on with two small bright lobes and a slightly pinched waist.  Their resulting inclination range was $i\approx$10--15$^\circ$
for the nebula.  They use $d=1.5$ kpc from \citet{sab86} to determine a nebular age of $\sim8700$ years.
Their value is in agreement with that from \citet*{sta97} who give an evolutionary time for the CS of 7400 years along with
$M_{CS}=5.87$ M$_\odot$.  So Sp~1 appears to be roughly one-third the age of NGC~6337.

\citet{jon12} also describe photometric results demonstrating smooth variability with no sign of fast scatter due to
mass transfer.

\begin{figure}[p]
\begin{center}
\plottwo{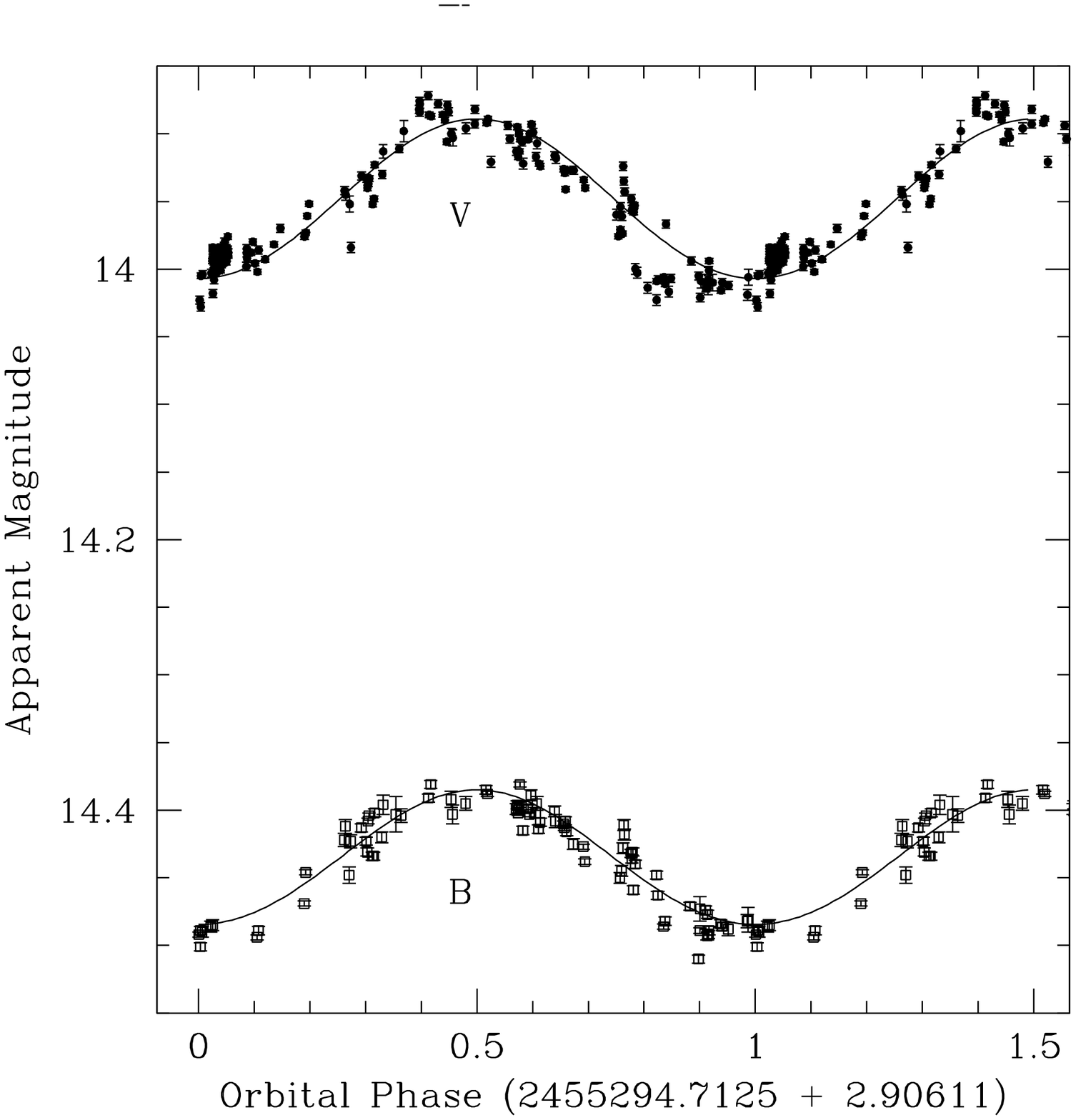}{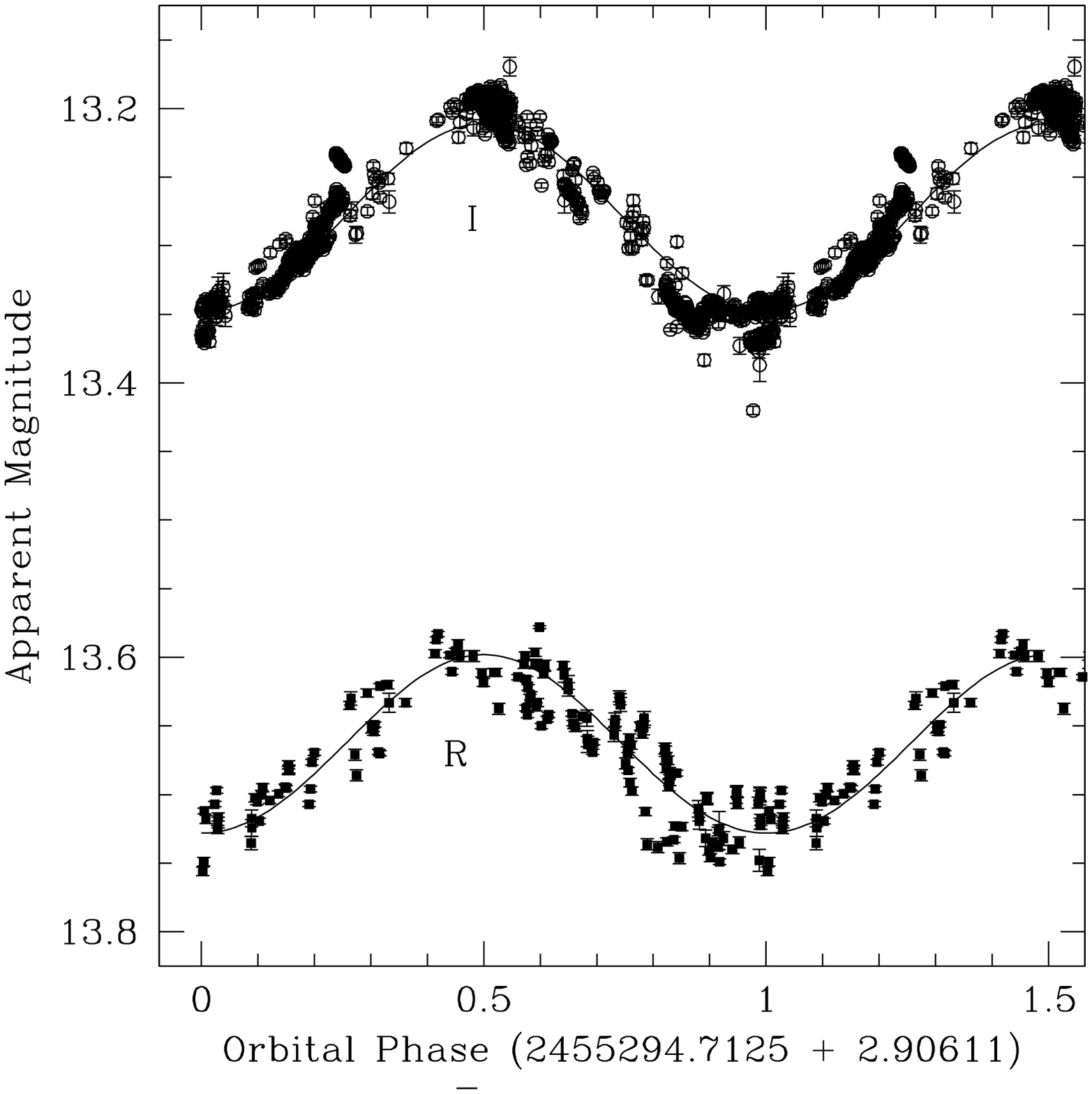}
\end{center}
\caption[Sp~1 Phase-Folded Light Curve] {The phase-folded $B$, $V$,
$R$, and $I$ light curves
for the CS of Sp~1 for the given ephemeris (Equation \ref{eq3}).  Also shown are the
model light curves from the Wilson-Devinney code described in \S 4.1.2.  
No reddening correction has been applied. \label{sp1phot}}
\end{figure}

\begin{figure}[p]
\begin{center}
\includegraphics[width=6in]{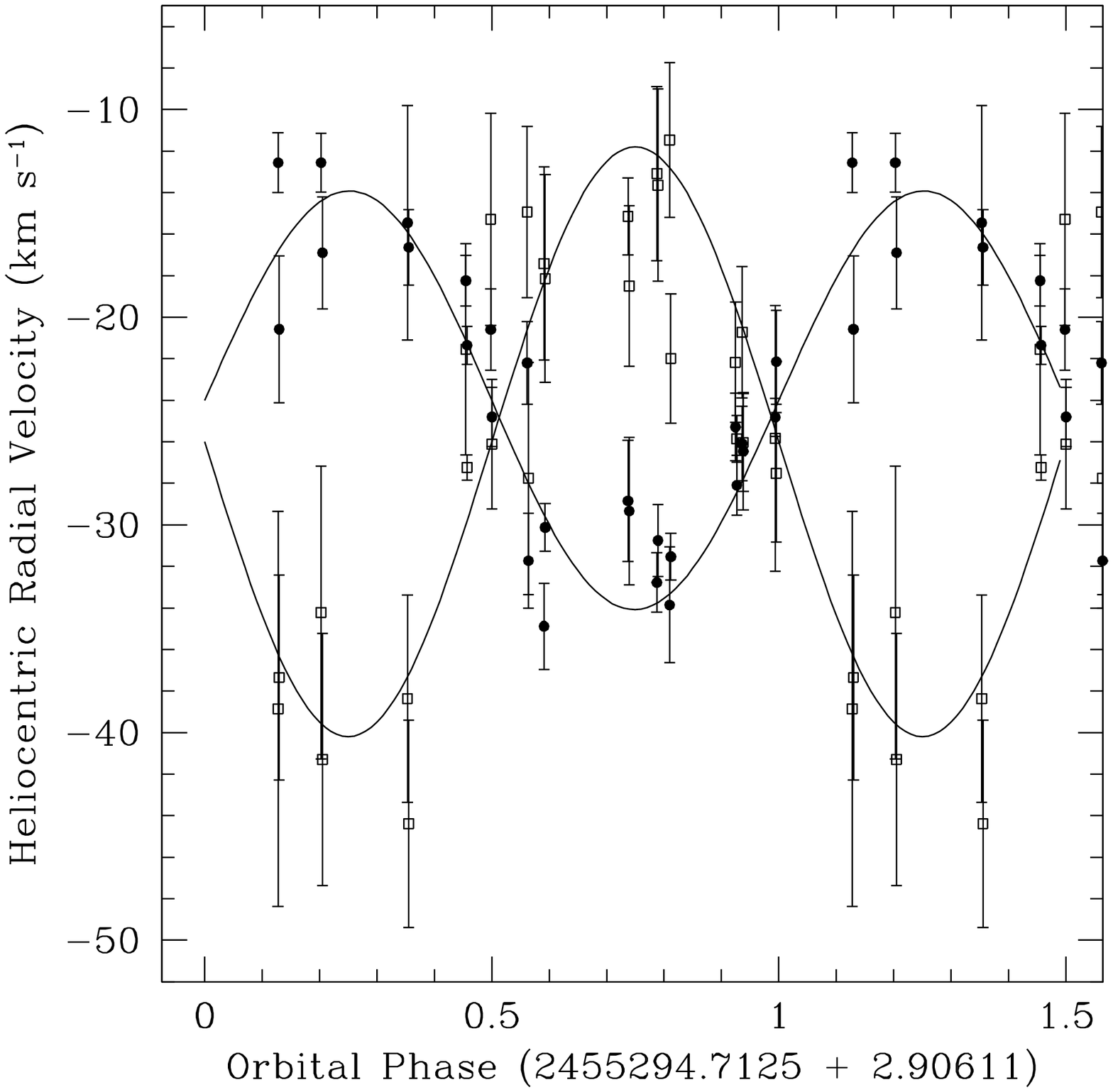}
\end{center}
\caption[Sp~1 Phase-Folded Radial-Velocity Curve] {The phase-folded radial-velocity curve
of the CS ({\it hollow squares}) and cool companion ({\it solid circles}) in Sp~1 for the period given in the ephemeris (Equation \ref{eq3}).  
The fit lines are from the Wilson-Devinney model described in \S 4.2.  We find a system
velocity of $\gamma=-24$ km s$^{-1}$ with a radial-velocity amplitude for the 
cool companion of $K_{2}=9.2\pm0.8$ km s$^{-1}$ and for the CS of $K_{CS}=13\pm2$ km s$^{-1}$.
\label{sp1rv}}
\end{figure}

Using the photometry described in \S 2.1 (Figure \ref{sp1phot} along with the radial velocity curve described in \S 2.2
(Figure \ref{sp1rv}), we find an updated orbital period in HJD of
\begin{equation}
T=2455294.7125(2)+2.90611(3)E
\label{eq3}
\end{equation}
for the binary CS of Sp~1.  As mentioned in \S 2.1 the period is consistent over the entire 23 year span of data.  However, 
the scatter about the sinusoidal irradiation effect is clearly larger than would be expected based on the photometric errors.  Our
results are consistent with the observation of \citet{jon12} showing no evidence of fast ``flickering'' that may be associated
with mass transfer or even an accretion disk in the system.  The scatter appears to be based on relatively slow variations in
brightness.  In order to explore this more, we removed the irradiation variability by subtracting our resulting binary model
(described below in \S 3.2.2) from the light curves and performing a period search on the resulting difference light curve
using the {\it Period04} software.  For the entire sample we find no consistent periodicity.  However, for data obtained over
relatively short times we find several potential periodicities showing nearly sinusoidal variability.  Unfortunately there is 
significant aliasing due to data sampling and none of the potential periods are consistent from one epoch of data to another.
At this point we are unable to determine the source of the scatter or if it is periodic in nature.

We note that in several filters the data appear too faint around phase 0.8-0.9, most notably the $V$ and $I$ bands.  It is possible
that this is a systematic trend in the light curve, potentially due to heat transfer away from the substellar point.  However, the
effect is less noticeable in the $B$ and $R$ light curves.  Most notably, the data in $V$ and $I$ at phase 0.8 were taken on the
same nights (along with some $R$ data) in 2009.  However, no $B$ data was taken at that time.  Additional $R$ data
taken in 2009 on different nights adds significant scatter at these phases, some of which is brighter than the model.  It is
possible that the faint appearance near phase 0.8-0.9 is simply due to a predominance of data taken on only a few nights
at a time when the system was in the fainter state of its slow brightness variations (non-irradiation related).

\subsection{Sp~1: Modeling the light and radial velocity curves}

Having light curves in four filters, along with radial velocity amplitudes for both stars in the binary system has
allowed us to determine the physical parameter values for the binary system with relatively tight constraints.
Inspecting the spectra we find that no \ion{He}{1} lines are visible, meaning a temperature of $> 70$ kK.
We find that a temperature of 80 kK matches the He II absorption lines relatively well, though with emission from
the nebula and irradiation lines from the companion filling in the lines, it is difficult to match them precisely.
However, the line wings suggest a $\log g$ value between 5.0 and 5.5.

For the purposes of our binary modeling we use $T_{CS}=80$ kK and try to maintain a $\log g$ value in the range
suggested above.  Using these two values and the mass determined by \citet{sta97} of $M_{CS}=0.587$ M$_\odot$
we find a luminosity that roughly agrees with post-AGB models for stars of that mass and luminosity.

From the irradiation lines in the spectrum we find a projected orbital velocity
relative to the center-of-mass for the cool companion of $K_{2}=9.2\pm0.8$ km s$^{-1}$ and a zero-point velocity
of $\gamma = -24\pm2$ km s$^{-1}$.  This is in agreement with the heliocentric systemic velocity of the nebula
determined from the nebular lines which we find to be $\gamma_{PN} = -27\pm2$ km s$^{-1}$.
Our values also agree with the value from \citet{mea88} of $\gamma_{PN} = -31\pm3$ km s$^{-1}$
but do not agree well with \citet{jon12} who find $\gamma_{PN} = -18\pm5$ km s$^{-1}$.

The radial velocity amplitude of the CS determined from the \ion{He}{2} absorption lines is found to be
$K_{CS}=13\pm2$ km s$^{-1}$ with a zero-point velocity
of $\gamma = -27\pm4$ km s$^{-1}$.  
We use the radial velocity amplitude values to restrict the system mass ratio in our modeling.
However, because the irradiation lines trace the irradiated hemisphere rather than the center of mass, the velocity
amplitude for the companion is likely larger than the value quoted above.  Using the Wilson-Devinney code, and taking
the resulting curve from those models, based on the center-of-light of the star, we find that the center-of-mass velocity
is approximately 2 km s$^{-1}$ greater in amplitude than the center-of-light curve.  We then adopt 
$K_{2}=11.2\pm0.8$ km s$^{-1}$ and $K_{CS}=13\pm2$ km s$^{-1}$.
From these we find the system mass ratio to be $q=1.2\pm0.2$.

Even though the system is not eclipsing the information we obtain from the radial velocity data combined with the
irradiation effect amplitudes limit the system inclination to the range $7^\circ \lesssim i \lesssim 11^\circ$.  The
remainder of the parameter ranges from the modeling are given in Table \ref{CSmodels}.

The resulting parameters for the companion show a star too big for a MS equivalent of that mass, as we typically
see in these systems.  However, the temperature, which is typically higher in close binary CSPNe companions is
lower in this case for most of the parameter combinations.  For the lower end of companion masses, the stellar
temperature is slightly above that expected for a MS star of that mass.  At these masses though the radius is
approaching twice the MS value.  So either the companion is too cool compared to what we would expect, or
it is significantly larger than expected, even for an irradiated companion.

The resulting $\log g$ range for the CS is slightly smaller in the model results relative to that from the spectrum.  And
we find a most likely range for the CS mass in a range typical of CSPNe.  We compared the CS values with
post-AGB evolutionary models from \citet{sch83} and found that over most of the range our masses from 
modeling are consistent with those results and give
a very rough age range consistent with the dynamical age of \citet{jon12}.  However, the lower end of the CS
mass range in our models produces ages inconsistent with the dynamical
age of the nebula.  If we require the CS to match the dynamical age of the PN,
we find $M_{CS} \gtrsim 0.55 M_\odot$.

The binary parameters also allow
us to calculate a distance to the nebula.  The total intrinsic system brightness for the models used with the
minimum apparent magnitudes from our light curves give a nebular distance in the range 1.0 to 1.5 kpc, which is
again consistent with published values, especially those of \citet{fre16}.

The differences between our results and those of \citet{bod12} are due to the addition of the radial velocity data,
which limits the results to the lower end of their inclination range.  Also allowing our models to have companions
with radii and temperatures that did not match their MS counterparts allowed fits to the light curves at lower
inclinations, which were also consistent with the radial velocity data.


\section{Inclinations of Known PNe with Binary Central Stars}

In the current published literature we find eight PNe with close binary CSs for which an inclination has been
determined for both the nebula and the central binary.  Those eight are 
Abell~63 \citep{mit07,afs08,bel94},
Hartl-Triton~4 \citep{tyn12,hil16}, 
NGC~6778 \citep{mis11,gue12}, 
Abell~41 \citep{jon10b,bru01,shi08}, 
Abell~65 \citep{huc13,hil15a}, 
Henize~2-428 \citep{rod01,san15}, 
and the two presented here, NGC~6337 \citep{gar09} and Sp~1 \citep{jon12}.

\begin{figure}[p]
\begin{center}
\includegraphics[angle=-90, width=7in]{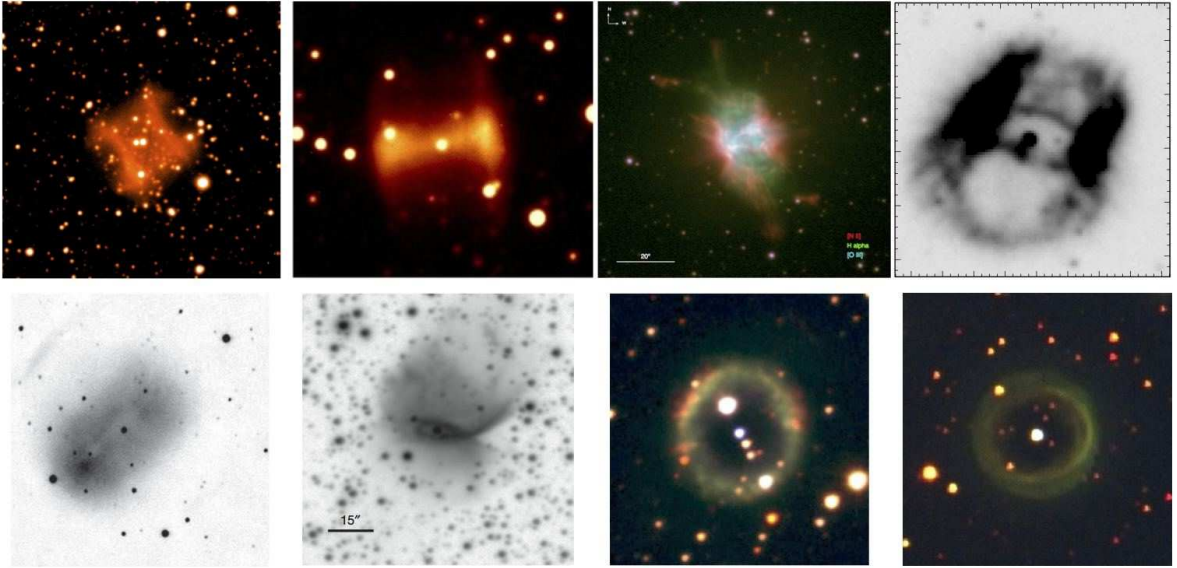}
\end{center}
\caption[PN Images] {Images of each of the eight PNe for which both nebular inclination and binary inclination for the
close binary CSs have been published.  {\it Top row}: Abell~63, HaTr~4, NGC~6778, and Abell~41.
{\it Bottom row}: Abell~65, Hen~2-428, NGC~6337, and Sp~1.  See the text for details.
\label{PNimages}}
\end{figure}

In Figure \ref{PNimages} we show images of each of these nebulae (not to the same scale).  Images are from
left to right: ({\it top row}) Abell~63 \citep{cor15}, HaTr~4 (ESO NTT archive), NGC~6778 \citep{gue12}, 
Abell~41 \citep{jon10b}, ({\it bottom row}) Abell~65 \citep{huc13}, Hen~2-428 \citep{san15}, NGC~6337 \citep{hil10},
and Sp~1 (our image).

In every case, the inclinations of the PN and central binary are consistent with one another, where {\it inclination}
for the PN is defined as the inclination of the primary axis of symmetry relative to the line-of-sight and for the binary
is the inclination of the orbital axis relative to the line-of-sight (as is typical for binary star analyses).
Therefore, agreement between the two inclinations means that, within the uncertainties, {\it the binary orbital axis is
aligned with the nebular symmetry axis}.  HaTr~4 is the only case in which the binary and nebular inclination ranges do not
overlap, though they do meet at the extreme and the binary inclination is based on modeling of the
light curves only and is thus a relatively uncertain range.

There is one potential example of a wide binary CSPN for which the binary and nebular axes do not align, the case of
LoTr~5.  \citet{van14} show that in order for the wide binary axis to align with the published value for the nebular
axis of $17^\circ$ \citep{gra04}, the hot CS would need to have a close companion with a mass of several solar masses.
However, there is currently no clear evidence for such a close companion.  This may then be a case of misalignment between
the binary and nebula.  The binary period is not known but is at least on the order of several thousand days, meaning
this system likely avoided the CE phase and is thus not analogous to the close-binary, post-CE systems.

\begin{figure}[p]
\begin{center}
\includegraphics[width=6in]{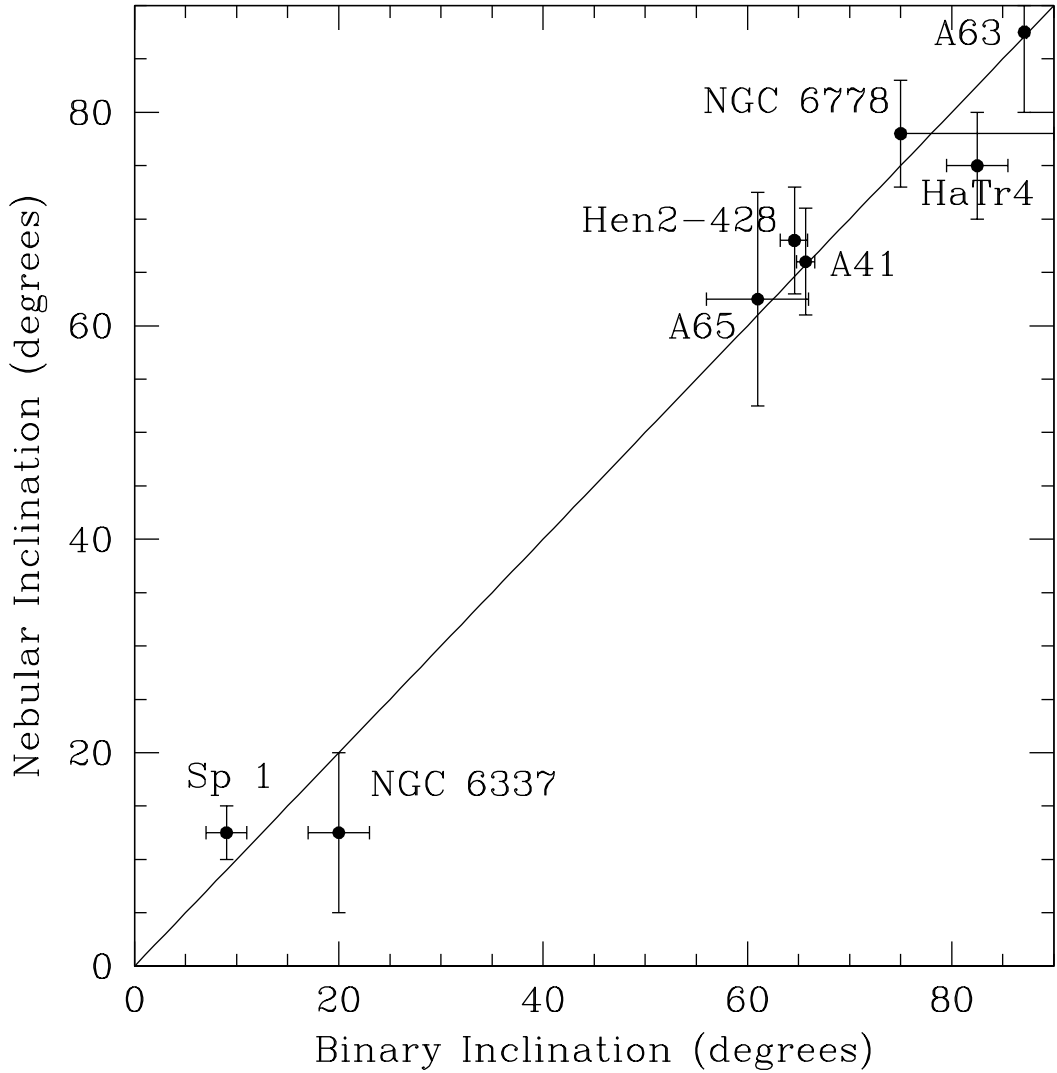}
\end{center}
\caption[Nebular and Binary Inclinations] {The nebular and binary system inclinations for
PNe with known binary CSs.  This represents all of the systems for which these values
have been modeled.  See the text for details.
\label{incl}}
\end{figure}

We show in Figure \ref{incl} the inclinations for each of the eight systems described above plotted against one another.
The solid line has a slope of one, showing where each data point would fall if the binary and PN are aligned.  The error bars for
the nebular inclinations are approximate values in those cases where the published description did not give
a range or error bars.  The binary inclination error bars are those quoted in the corresponding papers.

We have plotted the assumed value in the middle of the quoted range except for NGC~6778,
for which \citet{mis11} show the CS to be an eclipsing binary.  While no binary modeling has been published
for this object (we are pursuing formal modeling of the light curves and have recently obtained orbit-resolved
spectroscopy to measure radial velocity values), we take $75^\circ$ to be an approximate lower limit to the binary
inclination in order to see eclipses.  The light curve clearly shows partial eclipses, so it is unlikely that the binary
inclination is approaching $90^\circ$, but without further data we leave the top end of the range open.

While it may be tempting to perform a $\chi^2$ test on the data plotted in Figure \ref{incl} in order to statistically
explore the correlation between binary and nebular inclinations, there is an additional parameter not shown in the figure.
Like radial velocity, inclination does not tell us the three-dimensional alignment of the axis.  It is a projection of the
alignment onto the line-of-sight.  In each case we must also consider the PA of the axes.  We can typically determine
the PA of each PN from images and spectroscopy, but these values are not known for the central binaries.  
Thus the binary orbit may, or may not, be aligned with the nebula.  Statistically, a binary with inclination near
$90^\circ$ but with unknown PA still has a large uncertainty in the three-dimensional orientation of its orbital axis.  
However, because the binary inclinations for the CSs of Sp~1 and NGC~6337 are low, the PAs have a much
smaller effect on the possible range in physical orientation of their orbital axes.

To consider the physical alignment of a binary system, we take a sphere centered on the binary.  The inclination angle points
to some point on the sphere.  We can then rotate the system $360^\circ$ in PA about our line-of-sight.  The area swept out
on the sphere, given the uncertainty in the inclination, during that rotation describes the possible orientations of the axis.
That surface area on the sphere
relative to the area of a half sphere (since we do not discriminate between the two poles of the orbit) gives the fractional
area covered by the possible orbital orientation.  Thus for equivalent uncertainties a binary with low inclination, pointing
close to along the line-of-sight, will produce a much smaller fractional surface area.

If we determine the surface area on the same sphere for the
nebular inclination, then we can compare the two not only to see if they overlap, but also to determine the probability
that they would overlap if the two axial orientations were random and had no physical connection.  For random orientations,
taking the sum of the two fractional areas gives the maximum probability that they will overlap randomly.
For our purposes here we ignore the
PA measurements of the nebulae, thus giving an upper limit to the probabilities.  However as we show below even the
upper limit we find is convincing in terms of a physical relationship between PN shaping and the central binary.

The surface area on the sphere for a given inclination range from $i_1$ to $i_2$ and with no
PA value ($\int d\phi = 2\pi$) is given by
$$\int^{i_2}_{i_1} 2\pi r^2 \sin i di = -2\pi r^2 \cos i\mid^{i_2}_{i_1}.$$
With the surface area of a half-sphere equal to $2\pi r^2$ then the fractional area is just
$(\cos i_1 - \cos i_2)$.

Applying this equation to the inclination ranges published for the binary and nebular inclinations gives us the
results shown in Table \ref{stats}.  Column 6 in the table gives the sum of the fractional areas, $\sum$P, which
is the probability of a chance alignment of the PN and central binary.  The probabilities for NGC~6337 and Sp~1
are roughly 8\% and 3\% respectively.

\begin{deluxetable}{lccccc}
\tablewidth{0pc}
\tablecolumns{6}
\tablecaption{Inclination Ranges from the Literature and Probabilities of Agreement
\label{stats}}
\tablehead{
\colhead {PN}        & \colhead {$i_{bin}$}     & \colhead{$(\cos i_1 - \cos i_2)$ }  
& \colhead {$i_{neb}$}    &\colhead{$(\cos i_1 - \cos i_2)$}  & \colhead{$\sum$P}  \\
\colhead {Name}        & \colhead {(deg)}     & \colhead{(binary)}  
& \colhead {(deg)}    &\colhead{(nebula)}  & \colhead{ }  
}
\startdata
Abell~63      & 86.9--87.3  & 0.0070 & 80--90  & 0.1736  & 0.1806  \\
HaTr~4        & 80--85        & 0.1038  & 70--80  & 0.2549  & 0.3586  \\
NGC~6778  & 75--90        & 0.2588  & 73--83  & 0.1705  & 0.4293  \\
Abell~41      & 64.8--66.6  & 0.0286  & 61--71  & 0.1592  & 0.1879  \\
Abell~65      & 56--66        & 0.1525  & 52--72  & 0.3066  & 0.4591  \\
Hen 2-428   & 63.2--65.9  & 0.0425  & 63--73  & 0.1616  & 0.2042  \\
NGC~6337  & 17--23        & 0.0358  &  5--20   & 0.0565  & 0.0792  \\
Sp~1            & 7--11          & 0.0109  &  10--15 & 0.0189  & 0.0298  \\
\enddata
\tablecomments{Column 2 gives the binary system inclination from modeling, column 3 shows the fractional surface
area on a sphere covered by the binary inclination range, columns 4 and 5 give the same values but for the PN,
and column 5 gives the sum of the two fractional solid angles.}
\end{deluxetable}

While the individual probabilities of chance alignment are low for a number of these systems, the real significance
at this point is in the number of systems that we now have.  The probability that {\it all} of these systems
would be aligned by chance is the product of the individual probabilities.  Taking the product of the
$\sum$P values from Table \ref{stats} we find a 0.00009\% likelihood of random alignment of {\it all} of these
systems.

Does this require that the shape of the nebula is caused by the central binary?  No, not from this information alone.
These results demonstrate a very high correlation between the close binary and PNe axes.  For example, it is possible that
the individual stellar rotations are in the same plane as the binary orbit.  In that event, the stellar rotation axes
would also be aligned with the nebular axes and may have determined the nebular orientation.
It is possible that some large-scale effect, such as local or even galactic
magnetic fields determined the orientation of the binary when it formed {\it and} produced the shape of the binary
\citep[e.g. for a discussion of PNe and Galactic B-fields see][]{ree13}.

However, given recent work demonstrating the difficulty of having stellar rotation produce shaping, 
and the potential ability of binary systems to produce the observed shapes \citep*{nor07},
we believe that our results provide overwhelming observational evidence for a physical link between 
close binary CSPNe and the physical structure of their surrounding nebula.

\section{Discussion and Conclusion}

We have shown conclusively that the CSs of both NGC~6337 and Sp~1 are short period irradiated binary systems with
cool companions in low inclination orbits.  The companions in both systems appear to be larger than a main sequence
counterpart of the same mass, as is typical in these systems.  The companions are also typically hotter than expected
for a main sequence counterpart.  However, we find that in the case of Sp~1 the companion may be roughly the same
temperature, or cooler.  It is unclear why this is the case here.  The secondary {\it may} be slightly evolved, though there
is no evidence to support this possibility.  The companion in NGC~6337 does show a typically higher temperature.

The modeled inclinations of both binary systems also align with the inclinations of the surrounding nebulae.  And we show
that these two systems, now along with six other systems, make a sample of eight PNe for which the nebular inclination
and binary inclination of a close binary CS are known.  In all eight cases the two inclinations agree with one another,
within the uncertainties.  We demonstrate that the likelihood that all eight of these systems are aligned merely by
chance is vanishingly small.  All other known parameters considered, the conclusion of a causal link between binarity and
the axial symmetry of the PN is now on solid statistical grounds.

Post-CE PNe have already been tentatively associated with bipolar morphologies \citep{zij07,dem09,mis09b} although the link is not 
clear cut because some of the PNe do not show the distinctive bi-lobal structure \citep[e.g., M2-29;][]{haj08}. From CE simulations 
\citep[e.g.,][]{san98,pas12} it is clear that the ejection of the CE happens preferentially on the equatorial plane. 
This equatorial ejection is a torus with a very large scale height. The scale height is likely a function of the companion mass 
because it depends on the amount of angular momentum injected which is larger for a more massive companion.  A more 
massive companion also induces a stronger tide on the giant, which in turn results in more distant companions being captured 
into a CE interaction by the AGB star. These more massive and more distant companions carry more angular momentum into 
the envelope at the time of in-spiral.

Magnetic fields in CE interactions likely play an important dynamical role in the ejection \citep[e.g.,][]{reg95}, and can be investigated by jets observed in PNe. Some post-CE PNe have jets. Jet masses and kinematics have been measured \citep[e.g.,][]{jon14} and exploited to determine the likely magnetic field strength and geometry at the time of CE \citep{toc14}. They also allow us to determine the elusive timescales of common envelope events, such as the length of the pre-in-spiral phase or the formation of post-CE accretion disks after the in-spiral by a new phase of Roche lobe overflow or fall-back of envelope gas \citep{kur16}. 

Once the in-spiral takes place, it is over very quickly (of the order of a dynamical timescale of the giant which is between a month and a year for the giants of interest here). On the assumption that the envelope is actually ejected during the dynamical in-fall 
\citep[something that is currently debated;][]{nan15,iac16} we can assume that the orbital separation of the post-in-spiral giant is approximately the same as we see today. This means that the giant star is quickly changed to a star with a radius smaller than today's Roche lobe radius. Since the luminosity is fully dependent on the core mass, it does not change. As a result the temperature of the CS must incur a relatively similar increase. The least massive companions would, at least in principle sink deeper into the envelope of the giant to eject it. This may generate a correlation between orbital separations and companion mass, although one may have to account for more massive primaries likely needing more orbital shrinkage to eject the massive envelope. In conclusion, a number of complex correlations are expected between stellar, binary and PN parameters in these objects, something that we will be able to test thanks to continuous characterization of these objects increasing the number of accurate parameters available. Soon PNe will become the best testing ground of the common envelope interaction.

But for now we have demonstrated observationally a statistically significant connection between close binary CSs and their surrounding PN.

\acknowledgements
This material is based upon work supported by the National Science Foundation under Grant No. AST-1109683. Any opinions, findings, and conclusions or recommendations expressed in this material are those of the author(s) and do not necessarily reflect the views of the National Science Foundation
Based on observations (GS-2011A-Q-91) obtained at the Gemini Observatory, processed using the Gemini IRAF package, which is operated by the Association of Universities for Research in Astronomy, Inc., under a cooperative agreement with the NSF on behalf of the Gemini partnership: the National Science Foundation (United States), the National Research Council (Canada), CONICYT (Chile), Ministerio de Ciencia, Tecnología e Innovación Productiva (Argentina), and Ministério da Ciência, Tecnologia e Inovação (Brazil).
This paper uses observations made at the South African Astronomical Observatory (SAAO).

\end{document}